\documentclass[sigconf]{acmart}

\copyrightyear{2023}
\acmYear{2023}
\setcopyright{acmlicensed}\acmConference[ICMI '23]{INTERNATIONAL CONFERENCE ON MULTIMODAL INTERACTION}{October 9--13, 2023}{Paris, France}
\acmBooktitle{INTERNATIONAL CONFERENCE ON MULTIMODAL INTERACTION (ICMI '23), October 9--13, 2023, Paris, France}
\acmPrice{15.00}
\acmDOI{10.1145/3577190.3614203}
\acmISBN{979-8-4007-0055-2/23/10}

\usepackage{subcaption}
\usepackage{multicol}
\usepackage{nicefrac}
\usepackage{hyperref}
\usepackage{tabularx}
\usepackage[export]{adjustbox}
\begin{document}

\title[Predicting Player Engagement in \emph{Tom Clancy's The Division 2}]{Predicting Player Engagement in \emph{Tom Clancy's The Division 2}:\\ A Multimodal Approach via Pixels and Gamepad Actions}



\author{Kosmas Pinitas}

\affiliation{%
  \institution{Institute of Digital Games\\ University of Malta}\city{Msida}\country{Malta}
}
\email{kosmas.pinitas@um.edu.mt}

\author{David Renaudie}

\affiliation{%
	\institution{Massive Entertainment\\ Ubisoft}\city{Malmö}\country{Sweeden}
}
\email{david.renaudie@massive.se}

\author{Mike Thomsen}

\affiliation{%
	\institution{Massive Entertainment\\ Ubisoft}\city{Malmö}\country{Sweeden}
}
\email{mike.thomsen@massive.se}

\author{Matthew Barthet}

\affiliation{%
  \institution{Institute of Digital Games\\ University of Malta}\city{Msida}\country{Malta}
}
\email{matthew.barthet@um.edu.mt}

\author{Konstantinos Makantasis}

\affiliation{%
  \institution{Institute of Digital Games\\ University of Malta}\city{Msida}\country{Malta}
}
\email{konstantinos.makantasis@um.edu.mt}

\author{Antonios Liapis}

\affiliation{%
  \institution{Institute of Digital Games\\ University of Malta}\city{Msida}\country{Malta}
}
\email{antonios.liapis@um.edu.mt}

\author{Georgios N. Yannakakis}

\affiliation{%
  \institution{Institute of Digital Games\\ University of Malta}\city{Msida}\country{Malta}
}
\email{georgios.yannakakis@um.edu.mt}

\renewcommand{\shortauthors}{Pinitas et al.}

\begin{abstract}
This paper introduces a large scale multimodal corpus collected for the purpose of analysing and predicting player engagement in commercial-standard games. The corpus is solicited from 25 players of the action role-playing game \emph{Tom Clancy's The Division 2}, who annotated their level of engagement using a time-continuous annotation tool. The cleaned and processed corpus presented in this paper consists of nearly 20 hours of annotated gameplay videos accompanied by logged gamepad actions. We report preliminary results on predicting long-term player engagement based on in-game footage and game controller actions using Convolutional Neural Network architectures. Results obtained suggest we can predict the player engagement with up to $72\%$ accuracy on average ($88\%$ at best) when we fuse information from the game footage and the player's controller input. Our findings validate the hypothesis that long-term (i.e. 1 hour of play) engagement can be predicted efficiently solely from pixels and gamepad actions.    
\end{abstract}




\maketitle

\section{Introduction}


A largely unresolved challenge in the field of Affective Computing (AC) is the task of modelling affect over long periods of time. To examine the degree to which reliable long-term computational models of affect can be constructed, it is imperative to have access to corpora containing affect responses and annotations over extended time periods. The most widely used affect datasets \cite{melhart2022again,ringeval2013introducing,koelstra2011deap,kossaifi2019sewa}, however, contain sessions that last up to a few minutes, at most. 

Motivated by the lack of multimodal corpora for the study of long-term affect modelling, this paper introduces a game affect corpus consisting of 1-hour long interactive gameplay sessions. The introduced dataset contains data from 20 participants who played one hour of \emph{Tom Clancy's The Division 2} (Ubisoft, 2019)---\emph{The Division 2} for short---and annotated their own gameplay videos in terms of engagement using the PAGAN annotation tool \cite{PAGAN}. Apart from the in-game footage modality, the presented version of \emph{The Division 2} dataset also features the player's inputs on the game controller (gamepad). The long-term interactive nature of \emph{The Division 2} as an elicitor offers a unique contextual environment for modelling affect over extended periods of time, thereby broadening the research horizons of AC per se. The features of \emph{The Division 2} also capture the multimodal interaction capacities on popular and commercial-standard applications such as games.

To validate our hypothesis that in-game footage and gamepad actions can reliably predict player affect for extended periods of time, we train a number of Convolutional Neural Network (CNN) architectures on nearly 20 hours of frame and player input data of \emph{The Division 2} corpus. Inspired by \cite{makantasis2019pixels,makantasis2021pixels}, we assume that rich and sufficient affect information is existent (and interwoven) within the pixels and the actions of the gameplay, and thus that a predictive model will be able to capture it effectively. This initial study tests the efficiency of deep learning models (relying on these two modalities) on predicting long-term player engagement represented as a binary classification task (i.e. high vs. low engagement). Importantly for the nature of this study we also investigate the impact of time conditioning on the predictive capacity of long-term affect models. Our key results indicate that we can predict high and low engagement states from long-term affect stimuli with high accuracy, particularly when the fusion of the two modalities (game footage and gamepad actions) is time-conditioned.

This paper is novel in several ways. First, to the best of our knowledge, this is the first time a commercial-standard game such as \emph{The Division 2} is used for the study of long-term player experience and affect manifestations at large. Second, the paper presents a generic methodology for modelling affect by fusing pixel information with gamepad actions in this corpus. 
Third, we introduce the concept of \emph{time conditioning} for the purpose of modelling long-term engagement in games and beyond. Finally, the initial results presented in this paper serve as the baseline for this new multimodal corpus.

\section{Background} \label{sec:background}

This section reviews related work on affect and engagement modelling from pixels and other modalities (Sections \ref{sec:model_affect}- \ref{sec:model_engagement}) and moves on to survey literature on affect corpora (Section \ref{sec:affect_corpora}).

\subsection{Affect Modelling} \label{sec:model_affect}

Affective computing is a multidisciplinary field that studies the expression of emotions and aims to develop models that can computationally capture such manifestations \cite{picard2000affective}. As videos and images can elicit emotion, it comes as no surprise that affect modelling from visual cues is gaining ground. Before the advent of deep learning, the dominant approach involved the use of domain knowledge and high-level hand-crafted visual features \cite{zheng2010emotion,dahmane2011emotion}.
%
%
Although such approaches are memory efficient and allow for real-time emotion recognition, the development of large-scale affect datasets \cite{kossaifi2017afew,ringeval2013introducing} and the gradual advancement of deep learning led to significant breakthroughs in affect modelling \cite{martinez2013learning} and multimodal deep fusion \cite{martinez2014deep}. Indicatively, Breuer and Kimmer \cite{breuer2017deep} employed CNNs for various facial expression recognition tasks whereas Ng et al. \cite{ng2015deep} used CNNs pretrained on ImageNet \cite{russakovsky2015imagenet} to perform emotion recognition on small datasets. Assuming that games can be effective elicitors of affect, Makantasis et al. \cite{makantasis2019pixels} trained CNNs to map gameplay footage to arousal while Pinitas et al. \cite{pinitas2022rankneat} evolved parameters of a preference learner to predict arousal in gameplay videos.

Other modalities such as physiology and speech (audio) have also been  extensively used for modeling affect, either individually or in a multimodal setting \cite{abdullah2021multimodal,sebe2005multimodal,abbaschian2021deep,lieskovska2021review}. Notably, Martinez et al. \cite{martinez2013learning}
were the first to apply CNNs for detecting affect via physiological signals. Makantasis et al. \cite{makantasis2021pixels} employed CNNs and modelled arousal from raw gameplay footage and sound. Zhang et al. \cite{zhang2021deepvanet} used a Convolutional
LSTM  and a 1D-CNN to extract spatio-temporal facial and bio-sensing features, respectively. Recently, Pinitas et al. \cite{pinitas2022supervised} employed Supervised Contrastive Learning on audiovisual and physiological data to model arousal. Unlike the aforementioned studies, this paper presents preliminary findings regarding \textit{long-term} player engagement prediction from in-game footage and game controller input using CNNs. 

\subsection{Engagement Modelling}\label{sec:model_engagement}


 
It can be argued that \emph{engagement} plays an important role in human-computer interaction (HCI) as a multifaceted construct that encompasses cognitive, affective, and behaviourally characteristics of the user \cite{appleton2006measuring,bindl201032}. Given its pivotal role in HCI research, several studies have focused on modelling different aspects of user engagement. 
Dermouche and Pelachaud \cite{dermouche2019engagement} developed an LSTM-based model to predict user engagement in real time dyadic interactions based on facial expressions, head movements and gaze. Ting et al. \cite{ting2013student} employed Bayesian Networks to model variables of student
engagement in virtual learning environments, while Fan et al. \cite{fan2016robotic} presented a robotic coach system based on multi-user engagement.

Engagement modelling has been central to AC research because it facilitates the computational modelling of more complex emotional responses: different levels of engagement correspond to different arousal-valence points on the affective circumplex model \cite{russell1980circumplex} Indicatively, Vries et al. \cite{de2014data} propose a methodology for reverse engineering a consumer behaviour model for online customer engagement based on a computational and data-driven perspective. Games have also proven to be an engaging entertainment medium; consequently, it is unsurprising that there is a growing body of research in the field of player engagement modelling. Specifically, Melhart et al. \cite{melhart2020moment} used viewers' chat logs as a proxy for engagement and employed a small neural network to predict moment-to-moment gameplay engagement based solely on game telemetry. Xue et al. \cite{xue2017dynamic} proposed a Dynamic Difficulty Adjustment framework to maximise a player's engagement (as stay time). 
Finally, Huang et al. \cite{huang2019level} introduced a two-stage player engagement modeling approach using Hidden Markov Models.
In this paper we view engagement via the lens of affect (see Section \ref{sec:dataset-annotation}), and fuse captured gameplay footage and player actions to predict high or low engagement in different time segments of a long gameplay session. 



   
\subsection{Affect Corpora}
\label{sec:affect_corpora}

Over the years, affect modelling has relied increasingly on large-scale and data-hungry computational models, which in turn require extensive affect corpora that encompass quantifiable expressions of emotions elicited via appropriate stimuli. A commonly held view is that acquiring annotated data that contain reliable affect information is a fundamental aspect of this endeavour. As this study introduces and builds upon data from a large-scale affect corpus, this section provides an overview of the most commonly employed affect corpora and their characteristics.

A key distinguishing factor among affect datasets is the annotation protocol used. The \textit{first-person annotation} protocol involves participants performing a task and then annotating their own affect. For instance, MAHNOB-HCI \cite{soleymani2011multimodal} and DEAP \cite{koelstra2011deap} databases consist of multiple modalities, such as electroencephalogram, electrodermal activity, facial video, and others, recorded from a first-person perspective and annotated with affect labels via a first-person annotation protocol. However, growing body of work employs a \textit{third-person annotation} protocol, where participants perform a task while a team of annotators (usually experts) annotate the participant's emotions. Indicatively, the RECOLA database \cite{ringeval2013introducing} includes recordings of online dyadic interactions between participants solving a task in collaboration; a group of six experts provided the socio-affective data annotations at a later stage. A similar annotation protocol has been used in the SEWA database \cite{kossaifi2019sewa}, which consists of audio-visual recordings of participants discussing in pairs. While affect corpora in games are often based on first-person annotation of a recent playthrough \cite{melhart2022again}, Mavromoustakos et al. \cite{mavromoustakos2023hearthstone} tasked two external experts to annotate tension on a large corpus (over 26 hours of videos) of competitive Hearthstone matches.

Affect corpora tend to rely on audiovisual media such as music videos and movies \cite{zafeiriou2017aff,mollahosseini2017affectnet,kossaifi2017afew}. It is thus unsurprising that there is growing research attention on both board and video games \cite{doyran2021mumbai,mavromoustakos2023hearthstone,yannakakis2010towards} due to the fact that they are interactive elicitors offering rich affect information. One of the first video game-based affect corpora is the platformer experience dataset \cite{karpouzis2015platformer}, a collection of videos of \emph{Super Mario Bros} (Nintendo, 1985) players, facial cues, and gameplay features. The recent AGAIN dataset \cite{melhart2022again} offers game footage and event logs annotated for arousal in a continuous first-person fashion. The FUNii dataset \cite{beaudoin2019funii} features multiple recordings of electrocardiogram activity, electrodermal activity, controller input, gaze, and head position, and provides first-person annotations for fun, difficulty, workload, immersion, and user experience.

Unlike earlier multimodal affect corpora, this work analyses affect from a large-scale dataset of 20 gameplay sessions (nearly 1 hour each) annotated for engagement in a first-person manner. Moreover, we employ deep learning algorithms to model long-term engagement relying on users' multimodal signal streams. The dataset covered here contains only gamepad input and captured gameplay footage; however, the extended version of the dataset also features more participants and input modalities including electrodermal activity, photoplethysmography, and eye-tracking signals.

\section{Tom Clancy's The Division 2 Corpus} \label{sec:dataset}


This section presents the large-scale multimodal corpus of annotated gameplay videos for the action-role playing game \emph{Tom Clancy's The Division 2}. Section \ref{sec:dataset-game} describes the game, Section \ref{sec:dataset-collection} provides an overview of the data collection protocol and Section \ref{sec:dataset-preprocessing} covers our method for pre-processing the collected data.

\begin{figure}[!tb]
\centering
\includegraphics[width=\columnwidth]{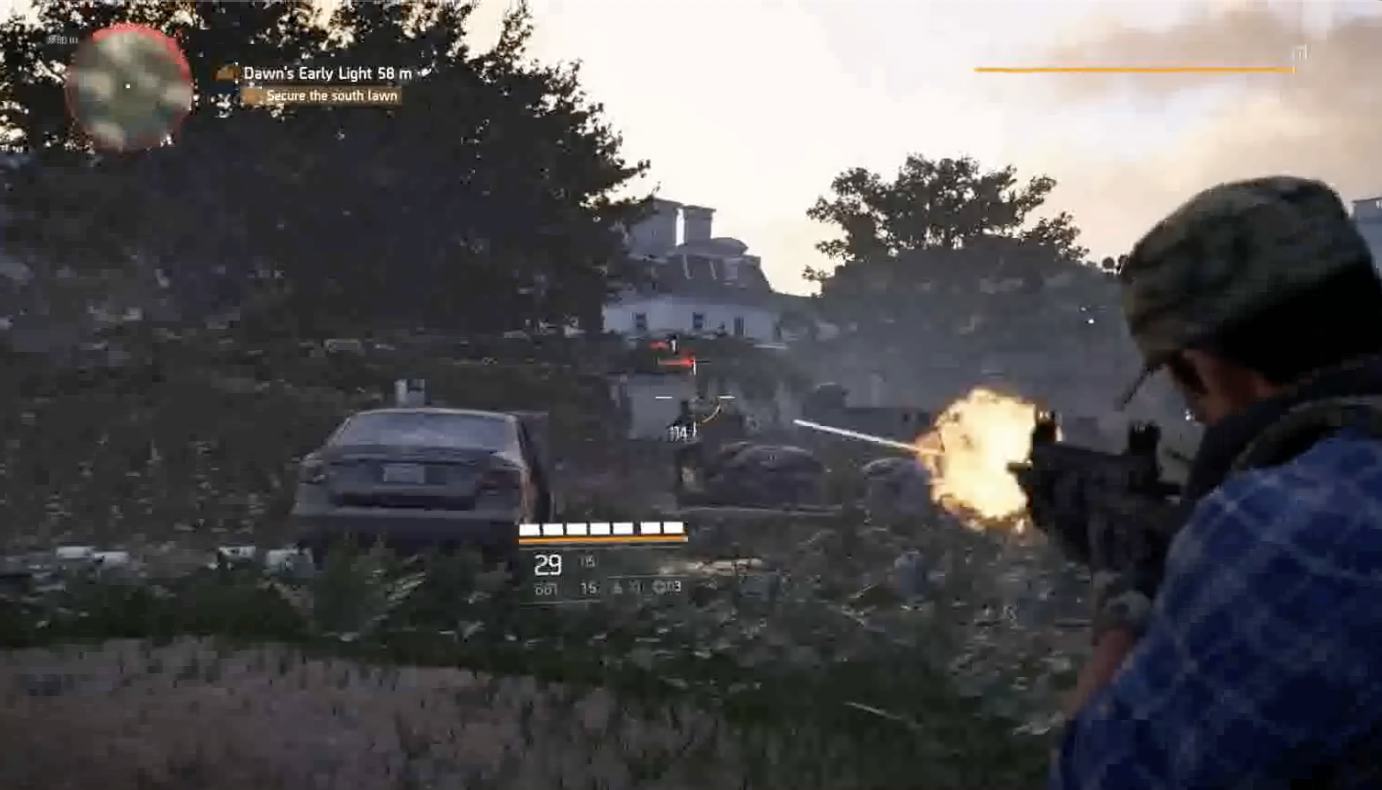}
\caption{In-game image of a player engaged in combat in \emph{The Division 2.}}
\label{fig:game}
\end{figure}

\subsection{The Game} \label{sec:dataset-game}

\emph{Tom Clancy's the Division 2} (Ubisoft, 2019) is an online action role-playing third-person shooter developed by Massive Entertainment and published by Ubisoft in 2019 (see Fig. \ref{fig:game}). The game, which has sold over 20 million copies worldwide, features both single-player and multi-player gameplay. Players can customise their characters and must scavenge for resources to survive in a challenging setting. The game contains over 30 missions where players must work their way through scripted content alone or in a group with other players. In this corpus, participants played the first mission of the game (\textit{Dawn's Early Light}), which offers a good balance between in-game exploration and combat. In this mission the player's goal is to stop the siege on the White House while securing the area. 

\emph{The Division 2} is a commercial-standard game environment which is ideal for eliciting rich affective responses. Besides high-quality graphics, \emph{The Division 2} has a complex input system and intense action set. Collectively, the properties of this game allow for a meaningful realisation of the affective loop \cite{yannakakis2014emotion}. The rich forms of HCI within the meticulously designed simulated world facilitate user immersion, which is essential for modelling the affective aspect of engagement. In the selected mission, stimuli are varied during a long gameplay session (approximately one hour). The balance between combat and exploration in this mission leads to intense user actions followed by periods of less intense activities.
 

\subsection{The Corpus}\label{sec:dataset-collection}

Data collection was carried out in two phases. First, 25 participants were asked to play roughly one hour of gameplay of \emph{The Division 2} using an XBOX controller (gamepad). All players played the same mission, \emph{Dawn's Early Light} (see Section \ref{sec:dataset-game}) alone (in single-player mode) until they completed the mission. 
Following the protocol introduced in \cite{melhart2022again}, participants were then 
asked to watch the recorded video of their own gameplay and annotate their \emph{engagement} in a continuous manner using the RankTrace \cite{ranktrace} annotation tool of the PAGAN platform \cite{PAGAN} (see Fig.~\ref{fig:game_hud}). At the beginning of the experiment, participants filled in a demographic survey. Building on ethical principles of AI and games research \cite{melhart2023ethics}, care was taken to ensure data was collected and analysed respecting GDPR principles. 

\subsubsection{Modalities of User Input}\label{sec:dataset-modalities}

\begin{figure}[tb]
\centering
\includegraphics[width=\columnwidth]{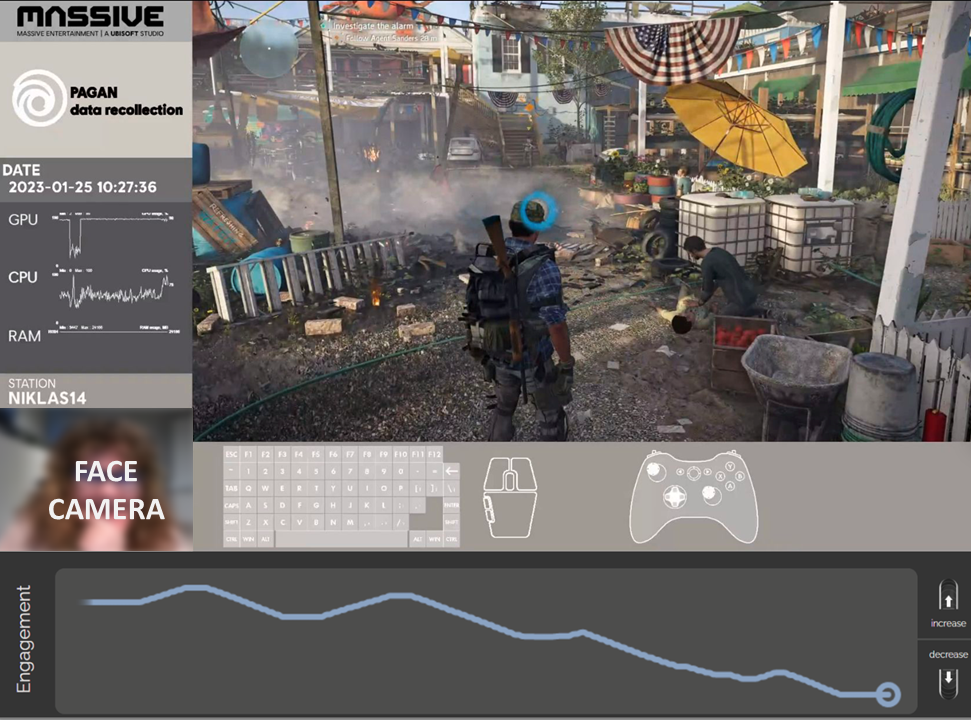}
\caption{Data collection snapshot for \emph{The Division 2}. Annotated videos contain timestamps for data synchronisation, statistics about compute resources, the ID of the workstation, the face of the participant (blurred out in this paper), eye-tracking data, and the live input on their gamepad. The bottom of the layout visualises the participant's engagement annotation trace using the PAGAN annotation tool.}
\label{fig:game_hud}
\end{figure}

During the gameplay phase of \emph{The Division 2}, several modalities were collected (see Fig.~\ref{fig:game_hud}). In this paper, we only process two types of information about the game context and the player behaviour. The \emph{frame} modality consists of a series of high-resolution frames of in-game footage  (usually 1280{$\times$}720 pixels). The \emph{gamepad} modality contains detailed player actions with the game controller. The possible gamepad actions captured in the dataset are 25, and include buttons pressed (e.g. ``A button pressed'') or other controller interactions (e.g. ``left stick up'') but do \textit{not} capture in-game events resulting from these inputs (e.g. ``player fires weapon''). These actions are stored in string format, and include co-occurring actions (combos). Frames and gamepad actions are stored and processed for our analysis (see Section \ref{sec:dataset-preprocessing}).

\subsubsection{Annotation}\label{sec:dataset-annotation}

After completing their gameplay sessions, participants were asked to annotate their experience in a first-person manner using the RankTrace \cite{ranktrace} annotation tool of the PAGAN platform \cite{PAGAN} (see Fig.~\ref{fig:game_hud}). PAGAN allows users to annotate a singular dimension in a time-continuous fashion while watching a video (in this case, the recorded gameplay data and other modalities as shown in Fig.~\ref{fig:game_hud}). PAGAN traces are synchronised to the video, and the annotators uses the mouse scroll wheel to increase or decrease the intensity of the perceived affect dimension. Resulting traces are unbounded, and thus annotators can keep increasing or decreasing this value as new stimuli are shown. The web interface of PAGAN shows the entirety of the trace so far, scaling the $x$-axis of the shown trace (Fig.~\ref{fig:game_hud}) as the video progresses.

Unlike the majority of continuously annotated affect corpora (see Section \ref{sec:affect_corpora}), this corpus tasked participants to annotate player experience with \textit{engagement} traces, instead of e.g. arousal or valence \cite{russell1980circumplex}. Participants were given the following definition of engagement prior to their annotation task: ``\emph{Engagement refers to the level of attention; a high level of engagement is associated with a feeling of tension, excitement, and readiness while a low level of engagement is associated with boredom, low interest, and disassociation with the game.}'' Although this definition of engagement closely resembles the definition of arousal, the latter refers to physiological activation and, consequently, it can not fully capture the complex and multifaceted nature of the gaming experience \cite{ledoux1998emotional,picard2000affective}. Additionally, engagement also captures aspects of valence since it can be a result of both positive and negative emotions such as happiness, terror, and anger \cite{latulipe2011love}. 
Based on the circumplex model of affect, we can argue that high engagement maps to high valence and high arousal while low engagement represents emotional states that are closer to the low arousal low valence quadrant of affect \cite{bindl201032}. Ultimately, we selected the annotation label of \emph{engagement} as it is highly representative of the gameplay context provided to our annotators whilst being related to the core affect dimensions of arousal and valence. 

To maximize the reliability and consistency of engagement annotations, participants were required to watch their gameplay at double speed (i.e. videos of 30 minutes) to minimize any effects caused by long-term annotation fatigue. The annotation trace is rescaled to the video duration before processing (see Section \ref{sec:dataset-preprocessing}).

\begin{table}[!tb]
	\centering
	\caption{\emph{The Division 2} Corpus Properties}
	\label{tab:dataset_properties}
	\begin{tabular}{|l|c|c|}
		\hline 
		Property & Raw & Clean \\
		\hline \hline
		Number of Participants & 25 & 20 \\
		\hline
		Number of Gameplay Videos & 25 & 20 \\
		\hline
		Number of Gamepad logs & 25 & 20 \\
		\hline
		Number of Annotated Video logs & 24 & 20 \\
		\hline
		Video database size & 24 hours & 18.8 hours \\
		\hline
		Number of Elicitors & \multicolumn{2}{c|}{1 game} \\
		\hline
		Gameplay video duration & \multicolumn{2}{c|}{53 to 65 minutes} \\
		\hline
		Annotation Perspective & \multicolumn{2}{c|}{First-person} \\ 
		\hline
		Annotation Type & \multicolumn{2}{c|}{Continuous unbounded} \\
		\hline
		Affect Labels &\multicolumn{2}{c|}{Engagement} \\
		\hline 
	\end{tabular}
\end{table}

\subsubsection{Dataset and Participants}

The raw dataset consists of 24 hours of gameplay (57.65 minutes per participant). However, only data from 20 participants are included in the clean dataset used here, due to inconsistencies on the annotation timestamps and missing data. This first iteration of \emph{The Division 2} engagement corpus includes the participants' controller inputs and gameplay frames (see Section \ref{sec:dataset-modalities}). Following \cite{melhart2022again}, we summarise the properties of \emph{The Division 2} corpus on Table \ref{tab:dataset_properties}.

Participants' ages ranged between 18 and 35, forming a diverse mix of individuals within the young adult category. 
Geographically, all participants are residents of Malm\"{o}, providing a localized perspective on the data. To collect precise data from electrodermal activity and eye-tracking, participants must not suffer from any skin condition or astigmatism. In terms of gaming experience, we aimed for participants that have not played the game before to ensure that they approached the study with a fresh perspective. However, a certain level of familiarity with the primary input method (XBOX controller) and other shooter games was required to acquire gameplay data of high and comparable quality.  



\subsection{Data Pre-Processing} \label{sec:dataset-preprocessing}

As this study aims to model long-term engagement via players' multimodal signals, we consider the following data pre-processing method. We split each participant's session (video) into overlapping time windows \cite{makantasis2022invariant,pinitas2022supervised} using a sliding step of 1.5 seconds and a window length of 10 seconds, corresponding to $22,541$ samples in the entire clean dataset. The sliding step and window length are essential hyperparameters since they influence, respectively, the size of the dataset and the information contained in each window.
The stimuli-based time windows (frames or gamepad modalities) are shifted by 1 sec to the annotation time window, accounting for the reaction time between stimulus and emotional response and the speed difference between gameplay and annotation \cite{pinitas2022rankneat}. 

After splitting each session into time windows, each window consists of a sequence of frames and logged gamepad actions. For the frame modality, we keep only 3 frames per second to reduce computational load. The 10 second time window used in this paper therefore consists of 30 RGB images of dimensions 224$\times$224$\times$3 (scaled down from the original high-resolution video). For the gamepad modality, we calculate the number of times the player pressed a specific key on the game controller during this time window, and also include a ``no key'' input as the number of times no key was pressed. Moreover, we calculate the number of $n$-button combos with $n$ ranging between 2 and 6. We convert these to input frequencies by dividing by the time window length. We thus collect 31 real-valued features from the gamepad modality (25 keypress frequencies, one ``no key'' frequency, and 5 combo frequencies), which are used as input to the model (see Section \ref{sec:methodology_arch}).

\begin{figure}[!tb]
\centering
\includegraphics[width=\columnwidth]{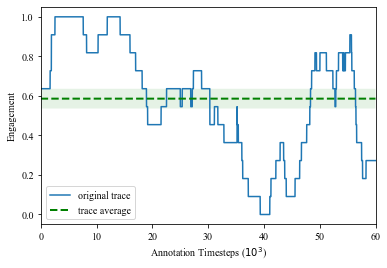}
\caption{The normalised engagement trace (blue) of participant $i$. The green line represents the mean value ($\mu_i$) of the trace and shaded area represents the ambiguity area $[\mu_i-\epsilon,\mu_i+\epsilon]$. Time windows with mean values above this shaded area correspond to high engagement; time windows below this shaded area correspond to low engagement.}\label{fig:engagement_trace}
\end{figure}

When it comes to the engagement traces, we perform a min-max normalization, transforming the unbounded engagement values to a value range of $[0,1]$ on a per-trace basis (see Fig.~\ref{fig:engagement_trace}). Similarly to the frames, we process the affect traces into time windows of 10 seconds. Finally, the average value of each time window provides a single engagement value per time window (see Section \ref{sec:methodology_learning}).


\section{Modelling Engagement} \label{sec:methodology}

We present our methodology for modelling engagement below. Section \ref{sec:methodology_learning} outlines the learning paradigm used to model engagement in \emph{The Division 2}, Section \ref{sec:methodology_arch} outlines the CNN architectures used for the unimodal (frames and gamepad actions) and multimodal network, while Section \ref{sec:methodology_time} presents different time-conditioning strategies explored.

\subsection{Learning Paradigms}
\label{sec:methodology_learning}

Arguably one of the most crucial steps in affect modelling is the choice of the supervised learning paradigm under which the mapping between multimodal user signals and affect labels will be inferred. When analysing an entirely new affect corpus such as \emph{The Division 2}, it is useful for all possible learning paradigms to be explored---including regression, classification and ordinal learning \cite{yannakakis2018ordinal}. In this initial study of \emph{The Division 2}, we focus on affect classification since it is one of the most commonly used learning paradigms in player and affect modelling \cite{makantasis2019pixels,makantasis2022invariant,makantasis2021privileged,pinitas2022supervised,makantasis2021pixels}.


In this first experiment, we follow the paradigm used in short-term time-continuous affect annotation traces \cite{makantasis2019pixels,makantasis2021pixels} and classify time windows based on the average trends of the entire 1-hour normalised trace (see Fig. \ref{fig:engagement_trace}). Specifically, we select classes based on the average affect value of the entire trace ($\mu_i$) of each participant ($i$) which acts as the class splitting criterion. For participant $i$, a time window $t$ is labelled as \textit{high engagement} when $e_{i,t} > \mu_i + \epsilon$ and as \emph{low engagement} when $e_{i,t} < \mu_i - \epsilon$; $e_{i,t}$ is the average normalised engagement value within the time window $t$ of participant $i$ (sampled at 30 Hz). It should be noted that the threshold $\epsilon$ is used to eliminate windows with ambiguous affect annotation values close to $\mu_i$, which may deteriorate the stability of the models. Following best practices from \cite{makantasis2019pixels,makantasis2021pixels} and preliminary tests with this corpus, we set $\epsilon=0.05$ for all experiments.

\begin{figure*}[!tb]
\centering
\subfloat[Gamepad model]{\includegraphics[width=.78\textwidth,left]{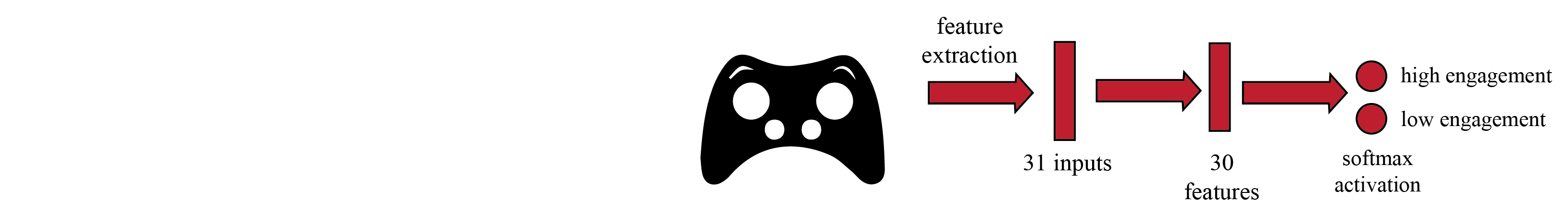}\label{fig:architecture1}}\\
\subfloat[Frames model]{\includegraphics[width=.8\textwidth,left]{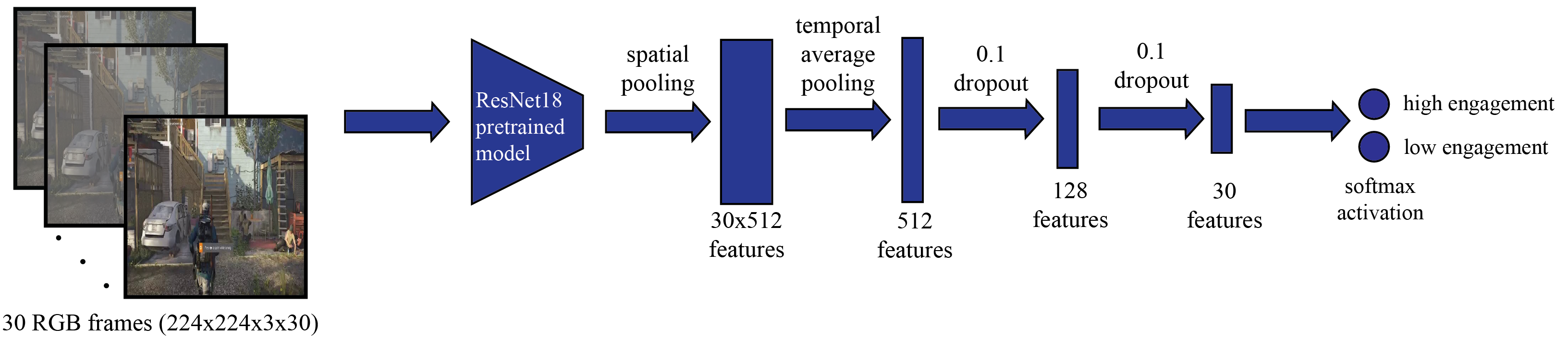}\label{fig:architecture2}}\\
\subfloat[Fusion model]{\includegraphics[width=.9\textwidth,left]{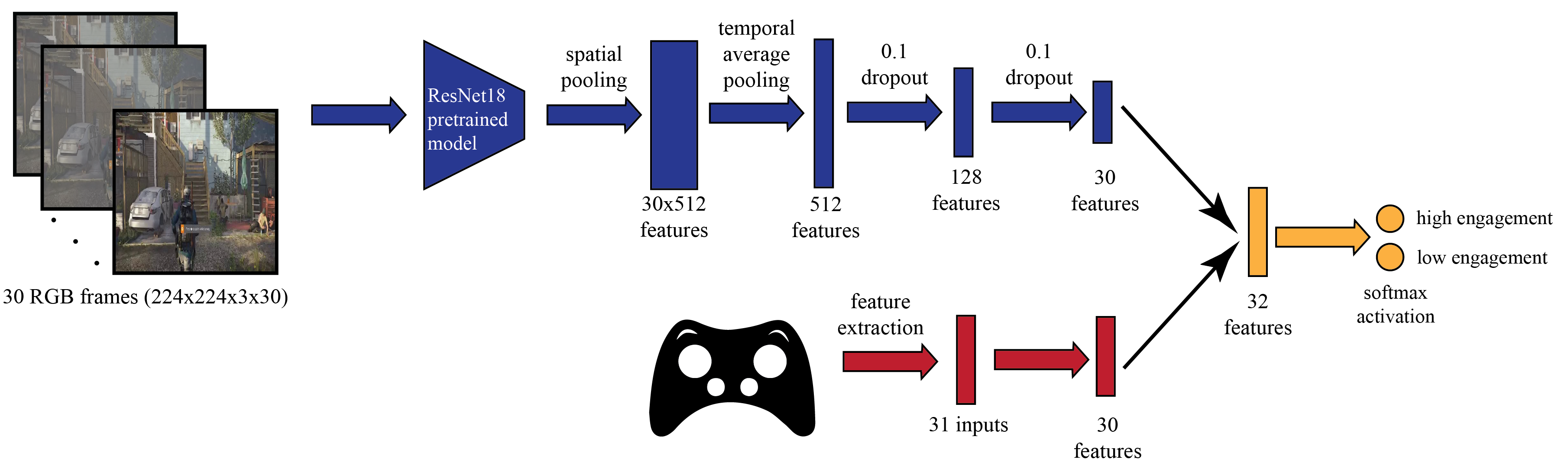}\label{fig:architecture3}}
\caption{The three model architectures employed for binary classification between high engagement and low engagement, using pixel information, gamepad actions, or both modalities through late fusion.}
\label{fig:architectures}
\end{figure*}

\subsection{Model Architecture} 
\label{sec:methodology_arch}

As mentioned in Section \ref{sec:dataset}, this paper considers two gameplay modalities of \emph{The Division 2}: (a) the player's controller input (i.e. \emph{gamepad} modality) and (b) the in-game footage (i.e. \emph{frame} modality). Since we treat the long-term engagement traces as a classification task (see Section \ref{sec:methodology_learning}), all architectures end with a 2-neuron softmax-activated layer that predicts low or high engagement.

The architecture used for the \emph{gamepad} modality is visualised in Figure~\ref{fig:architecture1}. A network takes the 31 inputs from gamepad actions (see Section \ref{sec:dataset-preprocessing}) and processes them via a Gelu-activated \cite{nguyen2021analysis} fully connected layer of 30 neurons, followed by a 2-neuron softmax-activated layer.

The architecture used to predict engagement from the \emph{frame} modality is visualised in Figure~\ref{fig:architecture2}. This network accepts 30 scaled-down RGB images as input (i.e. a tensor of $224\times224\times3\times30$ based on Section \ref{sec:dataset-preprocessing}) and processes them via a ResNet18 architecture that outputs $512$ feature maps of dimensions $7{\times}7$ per input frame, corresponding to a $30\times512\times 7\times7$ feature tensor. This ResNet18 architecture is pre-trained on ImageNet \cite{russakovsky2015imagenet}, similar to an abundance of previous work \cite{ng2015deep}, and its weights are frozen during this training process. The ResNet18 output passes through a spatial max pooling layer, reducing the dimensionality to $30 \times 512$, and a temporal average pooling returning a 1D vector of 512 features. The last vector is then fed into two consecutive Gelu-activated fully connected layers of 128 and 30 neurons respectively, each followed by a 0.1 dropout layer. Similar to the gamepad architecture, the last layer is a 2-neuron softmax-activated layer.

The fusion architecture considers both modalities (frames and gamepad actions) and is illustrated in Fig. \ref{fig:architecture3}. Following a \emph{late fusion} approach \cite{makantasis2021pixels}, the above unimodal architectures are combined by concatenating their latent representations to form a 1D vector of 60 features which is then fed into a Gelu-activated fully connected layer of 32 neurons followed by a 2-neuron softmax-activated layer.


\subsection{Conditioning on Time}
\label{sec:methodology_time}

Time conditioning refers to the practice of incorporating temporal information into machine learning models to improve their performance. When it comes to affect modelling, time conditioning can be of great value as the relationship between input variables (i.e. affect stimuli) and output variables (i.e. affect labels) changes over time---a phenomenon also known as \textit{concept drift} \cite{gama2014survey}. In order to validate our hypothesis that incorporating time features into an affect model can result in algorithms capable of capturing the dynamics of emotion over time, we explore three time-conditioning strategies that are detailed in the remainder of this section.

The first step of time conditioning involves transforming the scalar timestep value into a $D$-dimensional vector $e(t)$, which was first introduced by Transformers \cite{vaswani2017attention}. This vector offers a unique, deterministic and bounded encoding for each timestep while ensuring that the distance between any two timesteps is consistent across samples. The $e(t)$ vector is fed into a learnable linear down-projection layer that facilitates the injection of time in the models regardless of the size of the input modalities. Following preliminary tests, in this paper we use a vector of 512 features for all conditioning strategies, and treat the timestep at high granularity (increments of 20 minutes). The timestep $t_L$ can take three possible values depending on the time window's start time $t_w$, i.e. $t_L=1$ for $t_w\in[0,20)$ minutes, $t_L=2$ for $t_w\in[20,40)$ mins, and $t_L=3$ for $t_w\in[40,\infty)$ mins. The implementation of the 512-dimensional encoding is provided by the FAIRSEQ library \cite{ott2019fairseq} via Eq.~\eqref{eq:sinusoidal}:
\begin{equation}
	e(t) = \bigg[\ldots,\cos\big({t_L}{\cdot}c^{-\frac{2d}{D}}\big),\sin\big({t_L}{\cdot}c^{-\frac{2d}{D}}\big), \ldots \bigg]^T
 \label{eq:sinusoidal}
\end{equation}
\noindent
where $d = 1{\ldots}\nicefrac{D}{2}$ ($D = 512$ in this paper), $c = 10000$, and $t_L$ takes the values of 1, 2, or 3 depending on which 20-minute increment the time window belongs to.


\subsubsection{Shift Last Hidden Layer ($M_{SLL}$)} \label{sec:shift_last}
In this case, we follow the conditioning process employed in Decision Transformers \cite{chen2021decision}. The sinusoidal embedding is constructed via Eq.~\eqref{eq:sinusoidal} and is then down-projected linearly in order to match the dimensionality of the input of the model's last hidden layer. For the frames model (Fig.~\ref{fig:architecture2}), the last hidden layer is 30 features and $e(t)$ is down-projected to the previous layer (128 neurons). For a network with $H$ hidden layers the conditioned output is constructed as follows:
\begin{equation}
	O_{c,H-1} = O_{H-1} +s_{H-1}
\label{eq:sll}
\end{equation}
\noindent where $H$ is the last hidden (non-output) layer of the network, $O_{H-1}$ and $O_{c,H-1}$ is the output of the previous hidden layer ($H-1$), before and after conditioning, respectively; $s_{H-1}$ is the linear projection of the sinusoidal time embedding of the previous hidden layer.

\subsubsection{Scale and Shift Last Hidden Layer ($M_{SSLL}$)}
This method uses the same steps as in Section \ref{sec:shift_last}, but instead of learning a linear down-projection of size $n$ matching the dimensions of the penultimate hidden layer (e.g. $n=128$ in the frames model), we employ a linear projection of $2n$ neurons such that: 
\begin{equation}
	O_{c,H-1} = (l_{H-1}+1){\cdot}O_{H-1} +s_{H-1}
\label{eq:ssll}
\end{equation}
\noindent where $l_{H-1}$ and $s_{H-1}$ are, respectively, the first and last $n$ elements of the linear time embedding projection of the penultimate hidden layer ($H-1$); remaining notations are the same as in Eq.~\eqref{eq:sll}.

\subsubsection{Scale and Shift All Layers ($M_{SSAL}$)}\label{sec:scale_shift_everything}
Following \cite{voleti2022masked}, we explore the case of time-conditioning each layer as long as it is one-dimensional: e.g. in the frame model (Fig.~\ref{fig:architecture2}) the layers with 512, 126, 30, and 2 neurons are scaled and shifted. For each  layer $i$ of $n$ neurons, we learn a linear projection of $2n$ neurons such that:
\begin{equation}
	O_{c,i-1} = (l_{i}+1){\cdot}O_{i-1} +s_{i}
 \label{eq:ssal}
\end{equation}
\noindent where $l_{i}$ and $s_i$ are, respectively, the first and last $n$ elements of the linear time embedding projection for layer $i$; $O_{i-1}$ and $O_{c,i-1}$ are the outputs of the (previous) $i-1$ layer, respectively, before and after conditioning.



\section{Results}
\label{sec:results}

This section first outlines the experimental protocol we use to evaluate the algorithms and then presents the key results of the initial round of experiments performed with \emph{The Division 2} corpus.

\subsection{Experimental Protocol} \label{sec:protocol}

We test the capacity of the proposed modelling approaches to predict engagement in \emph{The Division 2}. The model is trained to classify frames and/or gamepad inputs within a time window as low or high engagement. Models in this paper are trained via the Adam optimiser with learning rate of $0.005$ and batch size of $256$. Moreover, we ensure that the same training, validation and test data are used for all models, promoting a fair comparison.

To evaluate model performance, we use a leave-2-participants-out cross-validation method. This method is a variant of the popular leave-one-participant-out cross-validation method \cite{kearns1999leaveoneout}, where two participants are used for the test set and another two participants are used for the validation set (for the purposes of early stopping). Data for training originates from 16 players, ensuring that data in each set belong to different participants and thus resulting in non-overlapping datasets. The models are trained for 50 epochs, but stop training after 5 epochs without a validation metric improvement; in all experiments in this paper, the maximum number of epochs was never reached. We split the dataset of 20 participants into 10 sets (with all participants becoming part of the 2-participants test fold) and calculate classification metrics on the test set averaged from 10 training runs, one per set. We randomise participant order and initial network weights 4 times, thus ensuring different participant pairings in the test set each time, and report results averaged across these train/test setups (i.e. 40 folds).

We benchmark our models based on the traditional \textit{accuracy score}, as the selected thresholding criterion ($\epsilon = 0.05$) ensures that the dataset is somewhat balanced. A naive \textit{baseline} uses the majority class in the training set and predicts the same class in the test set with an average test accuracy of 51\% across all folds (with a 95\% confidence interval of 0.28\% and a best-fold accuracy of 51.9\%). Statistical significance, when reported, refers to two-tailed paired Wilcoxon Signed-Rank Test with $p<0.05$, where data is matched on the same 2-participants' test folds. When multiple comparisons are performed, the Bonferroni correction is applied \cite{dunn2012bonferroni}.

\begin{figure}[!tb]
\centering
\includegraphics[width=.9\columnwidth]
{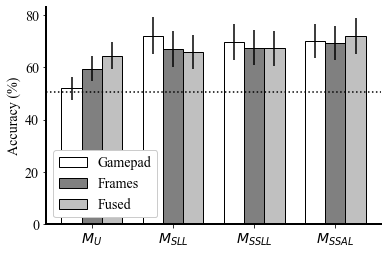}
\caption{Engagement classification in \emph{The Division 2}. The graph shows average (leave-2-participant-out) test accuracy values and corresponding 95\% confidence intervals. The test accuracy of the naive baseline is shown as a dotted line.}
\label{fig:accuracy}
\end{figure}

\subsection{Engagement Models without Time Context}\label{sec:experiment_baseline}

We report the average test accuracies from 10 cross-validation tests (via leave-2-participants-out) repeated 4 times in Figure \ref{fig:accuracy} under $M_U$. It is obvious that models trained on the gamepad modality alone perform poorly, with an average test accuracy of 52\% (best-fold accuracy of 68\%) which is very close to the baseline (no significant differences). Models trained on frames alone perform significantly better than the baseline and the gamepad models, with an average accuracy of 59.4\% (best-fold accuracy of 80\%). While gamepad data seems insufficient on their own, when fused with pixel information the trained models improve: the fusion model has an average accuracy of 64.5\% (best-fold accuracy of 82.4\%) which is significantly higher than all other models. As expected, late fusion of multiple modalities seems beneficial when it comes to engagement modelling, although accuracies remain low overall. This has been validated in previous work when fusing pixel and sound data in games \cite{makantasis2021pixels}, but not for controller input.

\subsection{Influence of Time Context}\label{sec:experiment_time}


The test accuracies of all three conditioning strategies described in Section \ref{sec:methodology_time} are shown in Fig.~\ref{fig:accuracy}, with unconditioned versions denoted as $M_U$. Surprisingly, the best accuracy for the gamepad modality is achieved with $M_{SLL}$ (average accuracy of 72\%). For the same conditioning strategy frames and fusion models perform significantly worse. A possible reason for this behaviour is the simpler architecture of the gamepad model (with only one hidden layer of 30 neurons), as conditioning applied only on the last hidden layer seems very effective. In comparison, scaling and shifting all hidden layers ($M_{SSAL}$) works better for the larger architectures, especially the fusion model which reaches accuracies of 72\% on average (best-fold accuracy of 87.7\%) and significantly outperforms the frames model on the same conditioning. Evidently, time conditioning is beneficial regardless of strategy applied: all conditioned models of gamepad or frames modalities perform significantly better than the unconditioned version for their respective modality.

\section{Discussion}\label{sec:discussion}

This paper introduced a novel and long-term player engagement multimodal corpus. The context of the multimodal interaction is the popular game \emph{Tom Clancy's The Division 2}. Findings of an initial engagement modelling experiment on this new dataset reveal that pixel information from the game footage can form efficient predictors of long-term player engagement. Without time embeddings, pixel information can be a strong predictor that is enhanced through fusion with gamepad actions to produce the best models, yet gamepad actions alone do not seem to be good predictors. One reason for poor gamepad models' performance could be the limited input size. Moreover, gamepad action logs lack the in-game context of the keypresses' effect on the game: for example, while gamepad inputs measure how often the player pressed the A button, the in-game effect of such an action may be very different depending on e.g. the avatar's currently held weapon. However, collecting in-game events in commercial games requires access to the game engine which may be unavailable due to intellectual property concerns. Therefore, the current experiment serves another purpose: to gauge to which degree data from player actions that respect current industry practices can be useful for affect modelling tasks.

The classification approaches presented in this initial study reveal that time embeddings are particularly efficient at capturing the long-term effects of player engagement with an average classification accuracy of $72\%$. We see three promising future research directions here. First, we plan to study and compare alternative learning paradigms such as preference learning which may capture informative local patterns---or changes \cite{yannakakis2017ordinal,yannakakis2018ordinal}---of engagement given the user modalities considered. Second, future studies will focus on different direct or indirect methods for integrating time within our multimodal models, including variants of LSTMs \cite{hochreiter1997long} and autoregressive models found in decision transformers \cite{chen2021decision}. Third, exploring other time embeddings with more granular time partitions (compared to the current 20-minute increments) may lead to breakthroughs in time conditioning for long-term affect prediction.

A core limitation of our first engagement modelling experiments is the baseline method we employ to derive the ground truth labels for the time windows of gameplay. Following current approaches in processing shorter gameplay sessions \cite{makantasis2019pixels,makantasis2021pixels}, we use the mean annotation value of a 1-hour trace to split windows into low or high engagement and leave ambiguous ones too close to the mean out of the train/test data. This approach is beneficial as it produces an almost equal split between class labels. However, summarising an entire 1-hour annotation session into one mean value overlooks possible habituation effects and the inherent subjectivity biases of human annotators \cite{yannakakis2018ordinal}, among many other factors. Furthermore, annotating lengthy audiovisual content can cause cognitive fatigue due to the mental exhaustion of the annotators \cite{souchet2022visual}, which in turn can affect the quality of the resulting trace. In future work, more nuanced ways of deriving classes should be explored, e.g., via a dynamically adjusted mean value derived from a moving time window of the trace. Initial experiments with a dynamic splitting criterion resulted in unbalanced datasets which in turn caused predictive models to underperform. Future work should explore signal processing approaches for deriving a more nuanced ground truth as well as improving algorithmic processes for modelling it. Another direction for future work that would address this issue is eschewing engagement classification altogether and treating consequent time windows in an \emph{ordinal} fashion \cite{yannakakis2018ordinal,yannakakis2017ordinal}. In such a treatment, the goal is to predict only whether the mean engagement between consequent time windows is (sufficiently) different, i.e. escalating or deescalating, which would discount for any long-term habituation or anchoring effects. We foresee several future directions for improving input data or engagement trace processing. 


While the full extent of \emph{The Division 2} corpus offers access to more modalities (including physiological signals and eye tracking), this initial study only focused on frame and gamepad action modalities. Inspired by earlier work \cite{makantasis2021pixels,makantasis2019pixels} we assume that in-game footage pixels combined with in-game actions would provide sufficient information for a model to predict player engagement accurately. While findings do corroborate our assumptions, including more modalities will likely improve the models' predictive power.

While \emph{The Division 2} dataset is not currently accessible, our short-term plan is to release the dataset and therefore encourage more research on the study of player engagement modelling via multimodal signals in a commercial-standard game environment.

\section{Conclusions}

The purpose of this paper is two-fold: (a) to introduce a novel dataset of long-term gameplay affect annotation traces that contains multiple modalities, and (b) to offer some initial suggestions and experiments on how such long-term affect traces can be processed and modelled. The extensive dataset analysed in this paper leverages two modalities---gameplay image frames and players' interaction data---but future work can explore more modalities already available in \emph{The Division 2} corpus. Experiments used a simple splitting criterion from the literature \cite{melhart2022again} to turn time-continuous annotations into binary classes, and demonstrated that gameplay frames can be good affect predictors as indicated in earlier studies \cite{makantasis2019pixels,makantasis2021pixels}. Our core findings suggest that long-term affect prediction is possible with high degrees of accuracy when time embeddings are injected to the model. The methods introduced here are generic and applicable to any study investigating long-term affect modelling.

\begin{acks}

This project has received funding from the European Union’s Horizon 2020 programme under grant agreement No 951911.
\end{acks}
\bibliographystyle{ACM-Reference-Format}
\bibliography{bibliography_camera_ready}

\end{document}